\documentclass[twocolumn,prb]{revtex4}
\usepackage{graphicx}
\usepackage{dcolumn}
\usepackage{bm}
\usepackage{amsmath}

\begin{document}

\preprint{}
\title{Resonant peak in the density of states in the normal metal / diffusive
ferromagnet / superconductor junctions}
\author{T. Yokoyama$^1$, Y. Tanaka$^1$ and A. A. Golubov$^2$}
\affiliation{$^1$Department of Applied Physics, Nagoya University, Nagoya, 464-8603, Japan%
\\
and CREST, Japan Science and Technology Corporation (JST) Nagoya, 464-8603,
Japan \\
$^2$ Faculty of Science and Technology, University of Twente, 7500 AE,
Enschede, The Netherlands}
\date{\today}

\begin{abstract}
The conditions for the formation of zero-energy peak in the density
of states (DOS) in the normal metal / insulator / diffusive
ferromagnet / insulator / $s$-wave superconductor (N/I/DF/I/S)
junctions are studied by solving the Usadel equations. The DOS of the DF is calculated in
various regimes for different magnitudes of the resistance,
Thouless energy and the exchange field of the DF, as well as for various
 resistances of the insulating barriers. The conditions
for the DOS peak are formulated for the cases of weak proximity
effect (large resistance of the DF/S interface)  and strong
proximity effect (small resistance of the DF/S interface).
\end{abstract}

\maketitle




%

%





In ferromagnet/superconductor (F/S) junctions Cooper pairs
penetrating into the F layer from the S layer have a nonzero
momentum due to the influence of exchange field
\cite{Buzdin1982,Buzdin1991,Demler}. This results in oscillating
behavior of the pair amplitude or a $\pi$-phase shift of the order
parameter in the ferromagnet. A negative sign of the real part of
the order parameter may occur when the thickness of the F layer is
larger than the coherence length of the F layer. The occurrence of
the $\pi $-phase shift makes it possible to realize the SFS $\pi $
-junctions \cite{Buzdin1982}, as was confirmed experimentally \cite%
{Ryazanov,Kontos1,Blum,Sellier,Strunk}. The order parameter
oscillations also lead to nonmonotonous dependence of $T_{c}$ in
SF bilayers on the F-layer thickness
\cite{Radovic,Tagirov,Fominov,Rusanov,Ryazanov1}. Effects of
resonant transmission in conductivity of SF structures were
discussed in Ref \cite{Kadigrobov,Leadbeater,Seviour}.


Another interesting consequence of the oscillations of the pair
amplitude is the spatially damped oscillating behavior of the
density of states (DOS) in a ferromagnet predicted theoretically
\cite{Buzdin,Baladie,Zareyan,Bergeret2} in various regimes. The
energy dependent DOS calculated in the clean \cite{Zareyan} and the
dirty \cite {Golubov} limits exhibits rich structures.
Experimentally DOS in F/S bilayers was measured by Kontos \textit{et
al}. who found a broad DOS peak around zero energy when the $\pi
$-phase shift occurs\cite{Kontos}. In diffusive
ferromagnet/superconductor (DF/S) junctions the zero-energy DOS may
have a sharp peak \cite{Golubov}. However the conditions
for the appearance of such anomaly have not been studied
systematically so far.

The purpose of the present paper is to calculate DOS in N/DF/S
junctions and to formulate the conditions for the zero-energy DOS
peak in two regimes corresponding to the weak proximity effect
(large DF/S interface resistance) and strong proximity effect (small
DF/S interface resistance). We will show that in the former case the
condition is equivalent to the one of Ref. \cite {Golubov}, while in
the latter case the new condition is found. The calculation will be
performed in the zero-temperature regime by varying the interface
resistances as well as the resistance, the exchange field and the
Thouless energy of the DF layer.


We consider a junction consisting of normal and superconducting reservoirs
connected by a quasi-one-dimensional diffusive ferromagnet conductor (DF)
with a resistance $R_{d}$ and  a length $L$ much larger than the mean free path. The DF/N interface
located at $x=0$ has the resistance $R_{b}^{\prime }$, while the DF/S
interface located at $x=L$ has the resistance $R_{b}$. We model infinitely
narrow insulating barriers by the delta function $U(x)=H\delta
(x-L)+H^{\prime }\delta (x)$. 
The resulting transparencies of the junctions $T_{m}$ and $T_{m}^{\prime }$
are given by $T_{m}=4\cos ^{2}\phi /(4\cos ^{2}\phi +Z^{2})$ and $%
T_{m}^{\prime }=4\cos ^{2}\phi /(4\cos ^{2}\phi +{Z^{\prime }}^{2})$, where $%
Z=2H/v_{F}$ and $Z^{\prime }=2H^{\prime }/v_{F}$ are dimensionless constants
and $\phi $ is the injection angle measured from the interface normal to the
junction and $v_{F}$ is Fermi velocity.

In the following calculation we will apply the quasiclassical
Green's functions formalism. The 2 $\times $ 2 retarded Green's
functions in N, DF and S are denoted by $\hat{R}_{0}(x)$,
$\hat{R}_{1}(x)$ and $\hat{R}_{2}(x)$
respectively. $\hat{R}_{0}(x)$ and $\hat{R}_{2}(x)$ are expressed by $ \hat{R%
}_{0}(x)=\hat{\tau}_{3} $  and $ \hat{R}_{2}(x)=(g\hat{\tau}_{3}+f\hat{\tau}%
_{2}) $ respectively, with $g=\varepsilon /\sqrt{\varepsilon
^{2}-\Delta^{2}}$ and $f=\Delta/ \sqrt{\Delta^{2}-\varepsilon
^{2}}$, where $\hat{\tau}_{2}$ and  $\hat{\tau}_{3}$ are the Pauli
matrices, and $\Delta$ and $\varepsilon$ denote the energy gap and
the quasiparticle energy measured from the Fermi energy,
respectively.
It is convenient to use the standard
$\theta$ -parametrization when function $\hat{R}_{1}(x)$ is
expressed as $ \hat{R}_{1}(x)=\hat{\tau}_{3}\cos \theta
(x)+\hat{\tau}_{2}\sin \theta (x). $ The parameter $\theta (x)$ is
a measure of the proximity effect in DF. The spatial dependence of $\theta (x)$  in DF is determined by
the static Usadel equation \cite{Usadel}
\begin{equation}
D\frac{\partial ^{2}}{\partial x^{2}}\theta (x)+2i(\varepsilon-(+)h )\sin
[\theta (x)]=0  \label{Usa1}
\end{equation}%
for majority (minority) spin 
with the diffusion constant $D$ and the exchange
field $h$ in DF. Note that we assume a weak ferromagnet and neglect the difference of
the Fermi velocities of the majority and minority spin subbands.


Further we shall apply the Nazarov's boundary condition
\cite{Nazarov2,TGK} for $\theta (x)$ at both interfaces. At the
DF/N interface it has the following form:
\begin{equation}
\frac{L}{R_{d}}\frac{\partial \theta (x)}{\partial x}\mid _{x=0_{+}}=\frac{
<F>^{\prime}}{R_{b}^{\prime}}  \label{Naza}
\end{equation}

\begin{equation*}
F = \frac{2T_{m}^{\prime} \sin\theta(0_{+}) } { (2-T_{m}^{\prime}) +
T_{m}^{\prime}\cos \theta(0_{+})}.
\end{equation*}

and it has a similar form at the DF/S interface. This boundary condition is based on the Zaitsev's boundary  condition\cite{Zaitsev} with  isotropic limit and generalizes the
Kupriyanov-Lukichev boundary condition \cite{KL}.

The average over the various angles of injected particles at the
interface is defined as
\begin{equation*}
<B(\phi)>^{\prime} = \frac{\int_{-\pi/2}^{\pi/2} d\phi \cos\phi B(\phi)}{
\int_{-\pi/2}^{\pi/2} d\phi T^{\prime}(\phi)\cos\phi}
\end{equation*}
with $B(\phi)=B$ and $T^{\prime}(\phi)=T_{m}^{\prime}$. The resistance of
the interface $R_{b}^{(\prime)}$ is given by
\begin{equation*}
R_{b}^{(\prime)}=R_{0}^{(\prime)} \frac{2} {\int_{-\pi/2}^{\pi/2} d\phi
T^{(\prime)}(\phi)\cos\phi}.
\end{equation*}
Here, for example,  $R_{b}^{(\prime)}$ denotes $R_{b}$ or $R_{b}^{\prime}$, and $R_{0}^{(\prime)}$ is Sharvin resistance, which in
three-dimensional case is given by
$R_{0}^{(\prime)-1}=e^{2}k_{F}^2S_c^{(\prime)}/(4\pi^{2} )$, where
$k_{F}$ is the Fermi wave-vector and $S_c^{(\prime)}$ is the
constriction area.


In the following, we will study the local DOS $N$ in the DF layer which is given by
\begin{equation*}
N/N_0 = \frac{1}{2}\sum _{ \uparrow , \downarrow } {\mathop{\rm Re}\nolimits}
\cos \theta(x)
\end{equation*}
where $N_0$ denotes the DOS in the normal state. The DOS will be
calculated by numerical solution of the Usadel equations with the
boundary conditions given above.




Below we will concentrate on the DOS at $x=0$ (N/DF interface) in
the regime of large resistance of the N/DF interface,
$R_d/R_b^{\prime} \ll 1$ and will also fix the barrier transparency
parameters $Z=3$, $Z^{\prime}=3$.

\begin{figure}[htb]
\begin{center}
\scalebox{0.4}{
\includegraphics[width=17.0cm,clip]{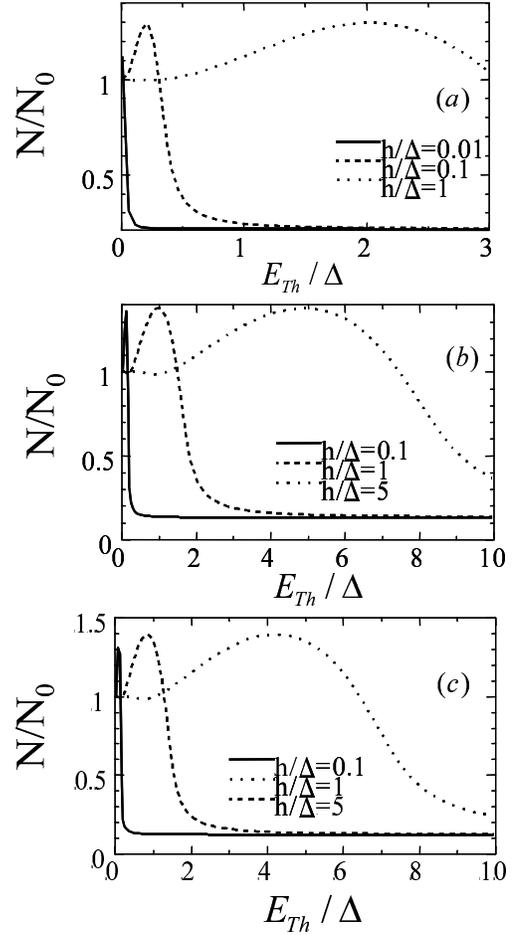}}
\end{center}
\caption{ Normalized zero energy DOS as a function of $E_{Th}$ for
large resistance of the N/DF interface $R_d/R_b^{\prime}=0.1$ and various $h/\Delta$ with resistance ratios at the DF/S interface (a)
$R_d/R_b = 1$, (b) $R_d/R_b =5$  and  (c) $R_d/R_b =10$.} \label{f1}
\end{figure}

In order to study the condition for the appearance of the zero energy DOS peak,
we plot the normalized zero energy DOS at $x=0$ as a function of
$E_{Th}=D/L^2$. Fig. \ref{f1} shows the DOS for
$R_d/R_b^{\prime}=0.1$ and various $h/\Delta$. In Fig. \ref{f1} (a)
the zero-energy peak appears at $E_{Th} \sim 2h R_b/R_d$, while in
Fig. \ref{f1} (b) and (c) the peak appears at $ E_{Th} \sim h$. Thus
we can conclude that the condition for the DOS peak for large
$R_d/R_b$ is essentially different from the one for small $R_d/R_b$.

\begin{figure}[htb]
\begin{center}
\scalebox{0.4}{
\includegraphics[width=17.0cm,clip]{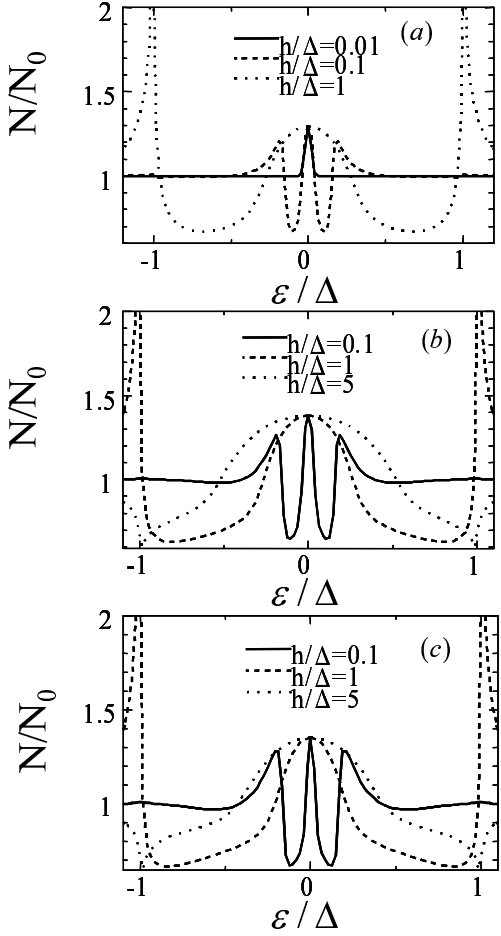}}
\end{center}
\caption{ Normalized DOS as a function of $\varepsilon$ for
$R_d/R_b^{\prime}=0.1$ and various $h/\Delta$ with (a) $R_d/R_b = 1$
and $E_{Th}= 2h R_b/R_d =2h$, (b) $R_d/R_b = 5$ and $E_{Th}= h$, and
(c) $R_d/R_b = 10$ and $E_{Th}= h$.} \label{f2}
\end{figure}

Fig. \ref{f2} shows the DOS as a function of $\varepsilon$ for the
parameters corresponding to the peaks in Fig. \ref{f1} for various
$h/\Delta$. In all these cases the DOS peak appears around zero
energy. For small $h/\Delta$ the DOS peak is narrow but it becomes
broader with the increase of $h/\Delta$. It's important to note that
this peak \textit{not} always requires the sign change of pair amplitude.
This is also clear from the fact that the peak occurs for large
Thouless energy (short DF) when there is no sign change. For other set of parameters the DOS peak is smeared as they break the condition $E_{Th} \sim 2h R_b/R_d$ or $ E_{Th} \sim h$.

\begin{figure}[htb]
\begin{center}
\scalebox{0.4}{
\includegraphics[width=17.0cm,clip]{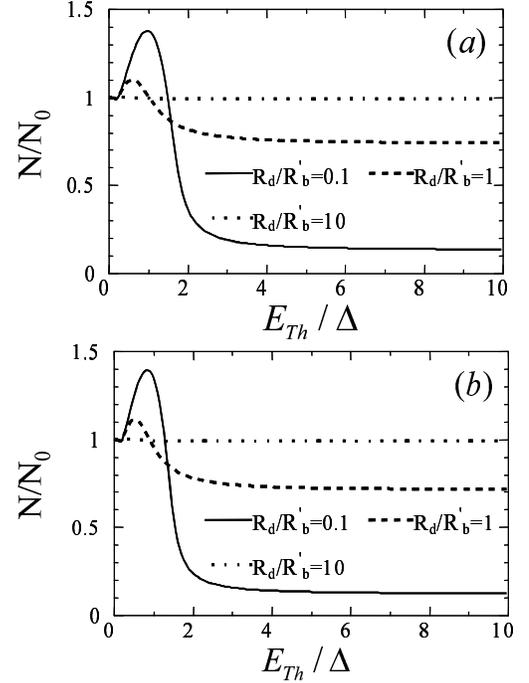}}
\end{center}
\caption{ Normalized DOS at zero energy as a function of $E_{Th}$
for $h/\Delta=1$ and various $R_{d}/R_{b}^{\prime }$ with (a)
$R_d/R_b = 5$ and (b) $R_d/R_b = 10$.} \label{f3}
\end{figure}

Let us first discuss the case of strong proximity effect  in
detail. Fig. \ref{f3} shows the zero energy DOS  at $x=0$ as a function of $E_{Th}$ for
$h/\Delta =1$ and various $R_{d}/R_{b}^{\prime }$ with (a) $R_{d}/R_{b}=5$ and  (b) $R_{d}/R_{b}=10$. In this
case the peak  at $E_{Th}\sim h$  is suppressed with increasing $R_{d}/R_{b}^{\prime }$. Therefore this
condition is valid for small $R_{d}/R_{b}^{\prime }$.

\begin{figure}[htb]
\begin{center}
\scalebox{0.4}{
\includegraphics[width=19.0cm,clip]{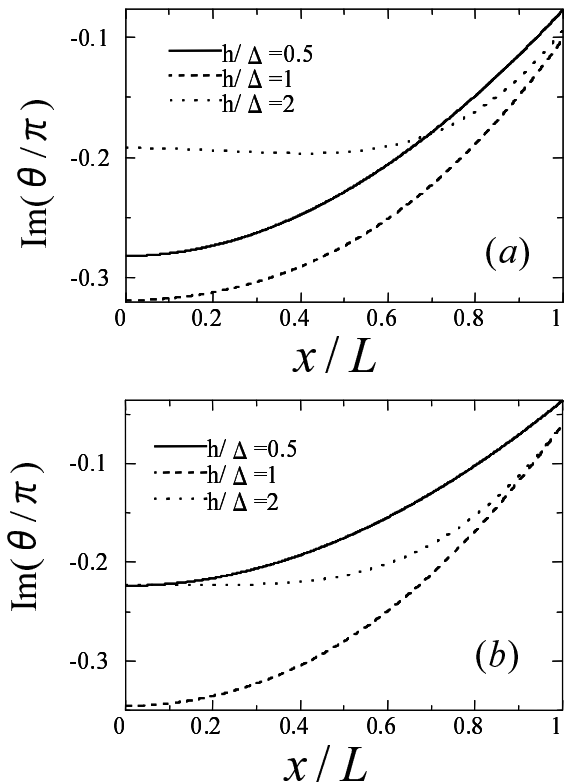}}
\end{center}
\caption{ Spatial dependence of Im$\protect\theta$ for majority spin
for $R_d/R_b^{\prime}=0.1$, $E_{Th}/\Delta =1$ and  various
$h/\Delta$ with (a) $R_d/R_b = 5$ and (b) $R_d/R_b = 10$. The DF/N
interface and the DF/S interface are located at $x=0$ and $x=L$
respectively.} \label{f4}
\end{figure}

Fig. \ref{f4} shows the spatial dependence of Im$\theta $ for
majority spin for $R_{d}/R_{b}^{\prime }=0.1$, $E_{Th}/\Delta =1$ and various $h/\Delta $ with (a) $R_{d}/R_{b}=5$ and (b)
$R_{d}/R_{b}=10$. For the appearance of the DOS peak, large value of Im$\theta$
is needed because the normalized DOS is given by Re$\cos (\theta
)=\cos (Re(\theta ))\cosh (Im(\theta ))$. As seen from Fig.
\ref{f4}, the magnitude of Im$\theta $ increases with the increase
of the distance from the DF/S interface and achieves a maximum
when $E_{Th}=h$.

Note that the zero-energy DOS at $x=0$ does not depend on $E_{Th}$ if the
condition $E_{Th}=h$ holds. To explain that, let's write Eqs. \ref{Usa1} and %
\ref{Naza} at $\varepsilon=0$:
\begin{equation*}
\frac{\partial ^{2}}{\partial y^{2}}\theta (y)-(+)2i\sin [\theta (y)]=0
\end{equation*}%
\begin{equation*}
\frac{1}{R_{d}}\frac{\partial \theta (y)}{\partial y}\mid _{y=0_{+}}=\frac{%
<F>^{\prime }}{R_{b}^{\prime }}
\end{equation*}%
where $y\equiv x/\sqrt{D/h}$. Since for $E_{Th}=h$ we have $D/h\equiv
E_{Th}L^{2}/h=L^{2}$,  the above equations don't contain $E_{Th}$ as a
parameter. Similar arguments can be applied to another boundary condition at
DF/S interface. This proves the above statement about independence of the
zero-energy DOS at $x=0$ on $E_{Th}$.

Now let us discuss the weak proximity effect and derive the condition $%
R_{d}/R_{b}\sim 2h/E_{Th}$, following Ref.
\cite{Golubov}. When spatial variation of $\theta $ is
small, i.e.,  $L \ll \sqrt{D/\mid \varepsilon \mp h \mid}$ (for the
spin-up or spin-down subband respectively) and both $R_{d}/R_{b}$ and $R_{d}/R_{b}^{\prime}$ are small (weak proximity effect),
$\theta $ can be
expanded as $\theta =\theta _{0}+\theta _{1}x+\theta _{2}x^{2}$ where $%
\theta _{1},\theta _{2}\ll \theta _{0}$. Note that the derivatives
of $\theta $ are proportional to these quantities at the interfaces
(see Eq. (\ref{Naza}) and Ref. \cite{TGK}).

In this case the solution of the Usadel equation in the spin-up
subband satisfying boundary conditions has the form:
\begin{equation}
\cos \theta _{0\uparrow }=\frac{{\frac{R_{d}}{R_{b}^{\prime }}+\frac{R_{d}}{%
R_{b}}g-\frac{{2i(\varepsilon -h)}}{E_{Th}}}}{{\sqrt{\left( {\frac{{R_{d}}}{{%
R_{b}}}f}\right) ^{2}+\left( {\frac{{R_{d}}}{{R_{b}^{\prime }}}+\frac{{R_{d}}%
}{{R_{b}}}g-\frac{{2i(\varepsilon -h)}}{{E_{Th}}}}\right) ^{2}}}}.
\end{equation}%
For $R_{d}/R_{b}^{\prime }=0$ and $\varepsilon =0$, the DOS has the form

\begin{equation}
\cos \theta _{0\uparrow }=\frac{{\frac{{2ih}}{{E_{Th}}}}}{{\sqrt{\left( {%
\frac{{R_{d}}}{{R_{b}}}}\right) ^{2}-\left( {\frac{{2h}}{{E_{Th}}}}\right)
^{2}}}},
\end{equation}%
which provides the resonant condition $R_{d}/R_{b}\sim 2h/E_{Th}$. Similar
result follows for the spin-down subband by replacing $h$ by $-h$.

Another resonant condition for the strong proximity effect, $E_{Th}\sim h$,
is equivalent to the condition $L\sim \sqrt{D/h}$. Thus, zero-energy DOS
peak appears when the proximity effect is strong and the length of
ferromagnet is of the order of the coherence length in a ferromagnet $\xi
_{F}=\sqrt{D/h}$. 

Let us discuss the physical meaning of two conditions. In DN/S junctions there is a minigap $E_g$, where $E_g \sim E_{Th} R_d/R_b$ for weak proximity effect,
 or 
$E_g \sim E_{Th}$ for strong proximity effect\cite{Golubov2}.
In DF/S junctions this minigap is shifted by $h$, then the DOS peak appears when $h \sim E_g$.

Note that in the calculations we have fixed $Z=Z^{\prime
}=3$, but the specific parameter choice does not change the results
qualitatively. 


In summary, we have studied the conditions for the appearance of the DOS
peak in diffusive ferromagnet, in normal metal / diffusive ferromagnet / $s$%
-wave superconductor junctions. We have discussed two regimes of weak and
strong proximity effect depending on the ratio $R_d/R_b$. The results in the
regime of weak proximity effect are essentially the same as found in Ref.
\cite{Golubov}. However, in the regime of strong proximity effect the
results are qualitatively different. Let us summarize the two conditions:

1. When the proximity effect is weak ($R_{d}/R_{b}\ll1$), the
condition for the DOS peak is $R_{d}/R_{b}\sim 2h/E_{Th}$.

2. When the proximity effect is strong ($R_{d}/R_{b}\gg1$), the
DOS peak appears when $E_{Th}\sim h$, i.e. when the length of
ferromagnet is of the order of the coherence length $\sqrt{D/h}$.

Note that the above two conditions cross over into each other when
$R_{d}/R_{b} \sim 2$. Since the DOS is a fundamental quantity affecting
various physical properties, our results may have many
applications, e.g., for the conductance of N/DF/S structures.


%
The authors appreciate useful and fruitful discussions with J.
Inoue, M. Yu. Kupriyanov, A. Kadigrobov, Yu. Nazarov and H. Itoh.
This work was supported by a Grant-in-Aid for the 21st Century COE
"Frontiers of Computational Science" and by INTAS Grant 01-0809.

%


\end{document}